\documentclass{jfm}

\usepackage{capt-of}

\usepackage{graphicx}
\usepackage{newtxtext}
\usepackage{newtxmath}
\usepackage{natbib}
\usepackage{hyperref}
\usepackage{nth}
\usepackage{caption} 
\hypersetup{
    colorlinks = true,
    urlcolor   = blue,
    citecolor  = black,
}

\newcommand{\RomanNumeralCaps}[1]
\linenumbers

\usepackage[x11names]{xcolor}       
\usepackage{soul}          

\newif\ifrevmode
\revmodetrue          

\ifrevmode
  
\else
  
\fi

\title{Diffusiophoretic dispersion of a colloidal blob in porous media}

\author{Aditya R. Pujari\aff{1} ,
 \and Amir A. Pahlavan\aff{1}\corresp{\email{amir.pahlavan@yale.edu}}}

\affiliation{\aff{1}Mechanical Engineering and Materials Science,\\
 Yale University, New Haven, Connecticut 06511, USA.}

\begin{document}
\maketitle

\begin{abstract}
Predicting and controlling the transport of colloids in porous media is essential for applications ranging from contaminant remediation to drug delivery. In these complex environments, solute gradients are ubiquitous and could drive diffusiophoretic particle migration, yet their impact on macroscopic colloid dispersion remains poorly understood. Here we combine experiments and simulations to quantify how diffusiophoresis alters the spreading of a colloidal blob in a 2D ordered/disordered porous medium. A joint blob of colloids and salt at high concentration is introduced into a medium filled with salt at low concentration and advected by a background flow. Intuition suggests that when colloids are attracted toward or repelled from the solute-rich blob, dispersion should be suppressed or enhanced, respectively. Instead, we observe the opposite trend: longitudinal dispersion is enhanced in the attractive case, whereas dispersion is suppressed in the repulsive case. Numerical simulations reveal that this striking reversal arises from diffusiophoretic exchange of particles between slow and fast streamlines, which we capture using a minimal two-layer model of coupled fast and slow plug flows. Finally, we probe how geometric disorder in the medium modulates this mechanism. Our results demonstrate that diffusiophoresis can strongly modulate macroscopic dispersion of colloids in porous media with implications for transport in subsurface and biological environments.
\end{abstract}



\section{Introduction}
\label{sec:headings}

Transport of colloids in porous media underpins a myriad of applications \citep{molnar_predicting_2015,bizmark_multiscale_2020,bordoloi_structure_2022}, from contaminant spreading and remediation in subsurface flows \citep{mccarthy_subsurface_1989}, oil and metal recovery \citep{park_microfluidic_2021, tan_two-step_2021}, and the spreading of microplastics and pathogens in groundwater \citep{pahlavan_soil_2024} to filtration \citep{florea2014long, kar2014particle} and drug delivery in tissues and biofilms \citep{ doan2021confinement, somasundar_diffusiophoretic_2023}. In many of these environments, chemical gradients are ubiquitous \citep{villermaux_mixing_2019,dentz_mixing_2023,borgne_fluid_2025} and can drive the motion of colloids via diffusiophoresis \citep{anderson1989colloid, shim2022diffusiophoresis, ault_physicochemical_2024}. This raises a natural question: how does pore-scale diffusiophoretic migration modulate the macroscopic transport and dispersion of colloids in complex environments?

The influence of solute gradients on colloid transport has been explored in a variety of settings, both without and with background flow. In the absence of flow, diffusiophoresis has been shown to drive particles into and out of dead-end pores \citep{kar_enhanced_2015, shin2016size, Alessio21, li2025solutedispersionboostsphoretic}. In the presence of flow, studies in channels, cellular flows, and advective systems have demonstrated that solute gradients can substantially reshape particle transport \citep{deseigne2014pinch, volk2014chaotic,raynal2018advection,chu2021macrotransport,  migacz2022diffusiophoresis,volk_phoresis_2022}. The impact of diffusiophoresis on transport in porous media has also been demonstrated, both without flow, e.g. in gels and biofilms \citep{doan2021confinement, somasundar_diffusiophoretic_2023, sambamoorthy2023diffusiophoresis}, and with flow, where phoretic migration to/from dead-end pores have been reported \citep{park_microfluidic_2021,jotkar_impact_2024}. Recently, we have shown that solute gradients in porous media can drive cross-streamline migration even within the main flow pathways, leading to substantial changes in colloid travel times and dispersion at the macroscale \citep{alipour2026}.

To gain insight into how diffusiophoresis and geometric disorder modulate macroscopic dispersion, here we investigate the spreading of colloidal blobs in a 2D lattice of ordered/disordered obstacles. We inject a blob of colloids suspended in a salt concentration $c_1$ into a porous medium initially filled with a different salt concentration $c_0$, and monitor the evolution of the blob under a background flow. We show that, depending on whether colloids move down or up the salt gradient, the longitudinal dispersion of the colloidal blob can be strongly suppressed or enhanced. We rationalize these observations using 2D numerical simulations and a minimal two-layer model. Finally, we demonstrate how geometric disorder modulates these effects.

\section{Microfluidic Experiments, 2D Simulations, and Two-Layer Model}
\begin{figure}
\centering 
    \includegraphics[width=\textwidth, keepaspectratio]{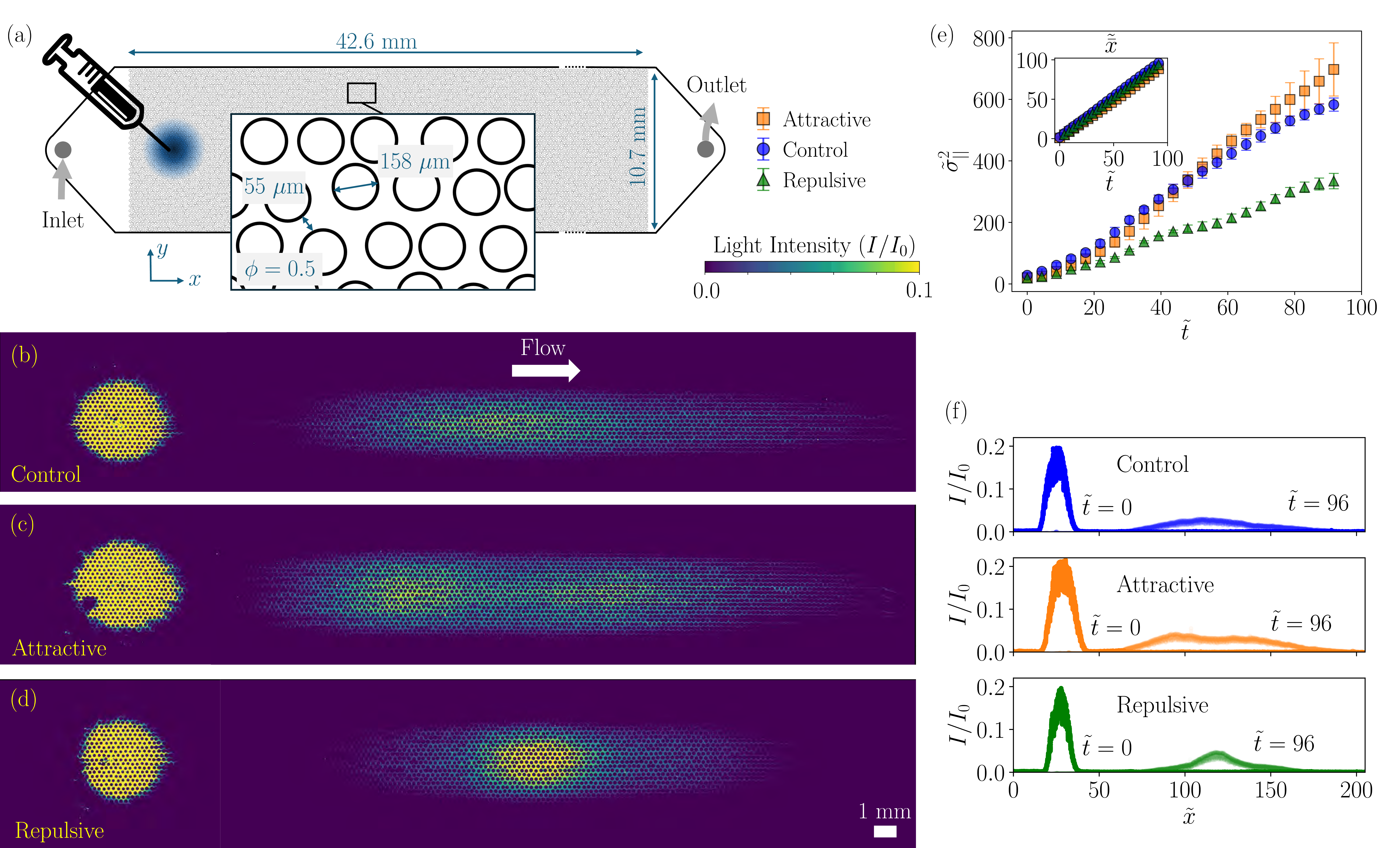}
    \captionsetup{width=\textwidth, justification=justified, singlelinecheck=false}  
\caption{\label{fig:first} Solute gradients significantly modulate the macroscopic dispersion of colloidal blobs in porous media. (a) Setup schematic showing the geometry of the porous medium consisting of a hexagonal lattice of circular posts with disorder strength $\chi=0.2$ and lattice spacing of $\lambda=213~\mu m$. (b,c,d) Two snapshots of the colloid blob at $\tilde{t}=0$ and $\tilde{t}=96$, where $\tilde{t}=tU_m/\lambda$. The colloidal blob in the attractive case splits into two blobs, while in the repulsive case the blob's dispersion is suppressed. Color scale is truncated for visual clarity. (e) Longitudinal dispersion $\tilde{\sigma}_{||}^{2}$ of the colloids as a function of time for 3 different cases: control, attractive and repulsive. Inset: The center of mass of all blobs $\tilde{\bar{x}}$ moves with the mean flow velocity. (f) The transverse-averaged light intensity corresponding to panels (b-d).}
\end{figure}

\noindent \textbf{Microfluidic Experiments:} The microfluidic chip is fabricated with standard soft photolithography \citep{xia1998soft} using Poly(dimethylsiloxane) (PDMS) and plasma-bonded to a glass slide (channel height $\approx 50 \mu m$). For each experiment, the porous medium chip (200x60 posts or 42.6x10.7mm) is filled with an aqueous solution of 0.01mM of LiCl or SDS using a syringe pump (Harvard Apparatus). Blob creation is done by piercing and injecting the microfluidic chip in situ with particles suspended in 1mM LiCl, or 0.01mM LiCl, or 1mM SDS for the attractive/control/repulsive cases, respectively (Fig.~\ref{fig:first}(a)). Flow begun $\sim10s$ after the introduction of the blob. Imaging is done using an inverted microscope (Nikon Eclipse Ti2 equipped with ORCA-Fusion Digital CMOS camera with a spatial resolution of 2300x2300 pixels and dynamic range 16 bit) and 200nm fluorescent particles (invitrogen), with excitation/emission wavelengths of 470/525 nm. We kept the particle concentration $n$ within the linear range of calibration correlating with pixel intensity. To image the entire length of the channel, we use large image stitching across 6 images at 0.2 fps.

The 2D porous medium is designed using circular posts of diameter $158 \mu m$ arranged in a hexagonal lattice with porosity $\sim 0.5$. To create disorder, each obstacle is displaced from its lattice position in a random direction $0<\Theta<2 \pi$ by a perturbation drawn from a uniform random distribution between $(0,\chi r]$, where $\chi$ is the disorder parameter and $r$ is the post radius. Any overlaps were removed by re-displacing the post with a new perturbation until there were no overlaps.
 
The solute ambipolar diffusivities ($2D_+D_-/(D_++D_-)$, where $D_+$ and $D_-$ are cation and anion diffusivities) are $D_{\textrm{LiCl}}=1.37\times 10^{-9} \textrm{m}^2/\textrm{s}$, $D_{\textrm{SDS}}=0.84\times 10^{-9} \textrm{m}^2/\textrm{s}$ and the particle diffusivity is $D_p=2\times 10^{-12}\textrm{m}^2/\textrm{s}$. Since the Debye layer ($\kappa^{-1}\approx 14 \textrm{nm}$ at 298K for 1mM Z:Z=1:1 electrolyte) is comparable to the particle radius, we use the finite Debye layer expression to calculate diffusiophoretic mobility \cite{lee2023role}: $\Gamma_p = \frac{\epsilon}{\eta} \left( \frac{k_BT}{Ze}\right)^2 \left[ \frac{Ze\zeta_p\beta}{k_BT}g(\lambda') + 4\ln \left[\cosh{\left(\frac{Ze\zeta_p}{k_BT}\right)}\right]h(\lambda')\right]$, where, $\epsilon$ is the permittivity of the medium, $\eta$ is the dynamic viscosity of the medium, $k_B$ is the Boltzmann constant, $T$ is the temperature, $Z$ is the valence of the solute, $e$ is the elementary charge, $\zeta_p$ is the zeta potential of colloidal particle, $\beta$ is the diffusivity difference equal to $(D_+-D_-)/(D_++D_-)$ for a Z:Z electrolyte, and $\lambda'=(\kappa a)^{-1}$ is the ratio of the Debye layer to the radius of the particle. $g(\lambda')$ and $h(\lambda')$ are functions given in \cite{lee2023role}. Although $\zeta_p$ may depend on the solute concentration, pH and the species of counter-ion, we assume a constant $\zeta_p\approx -70mV$ and therefore a constant $\Gamma_p$ in the simulations for simplicity \citep{gupta2020diffusiophoresis}. With this assumption, the corresponding constant diffusiophoretic mobilities are $\Gamma_{p,\textrm{LiCl}}=0.27\times 10^{-9} \textrm{m}^2/\textrm{s}$ and $\Gamma_{p,\textrm{SDS}}=-0.11 \times 10^{-9} \textrm{m}^2/\textrm{s}$. For the control case, we set $\Gamma_p=0$. \\

\noindent \textbf{Experimental Observations:} In the absence of solute gradients, the colloidal blob is advected by the background flow and becomes dispersed due to velocity heterogeneities (Fig.~\ref{fig:first}(b)). In the presence of solute gradients, colloids move up the gradient in LiCl case, where $\Gamma_p>0$, and down the gradient in the SDS case, where $\Gamma_p<0$. The mean longitudinal flow velocity is $U_m$ $\approx190~\mu m/s$ (equivalent to the mean migration velocity of the blob), corresponding to a $Pe_c = U_m\lambda/D_c \approx 1.8\times 10^4$ and $Pe_s= U_m\lambda/D_{\textrm{s}} \approx 30 \,(\text{LiCl, attractive}), 48\,(\text{SDS, repulsive})$. In the attractive case, the colloidal blob suspended in $1$ mM LiCl solution is injected into the medium with background salt concentration $0.01$ mM LiCl. We expect the colloids to remain more coherent than the control case as they are attracted toward the solute blob, suppressing the dispersion. Surprisingly, however, we observe the opposite: the colloidal blob becomes more dispersed in the longitudinal direction and even shows two peaks, indicating the splitting of the blob (Fig.~\ref{fig:first}(c)). By contrast, in the repulsive case, where the colloidal blob suspended in $1$ mM SDS solution is injected into the medium filled with $0.01$ mM SDS, we expect the colloidal blob to become more dispersed, and yet we observe a suppressed dispersion (Fig.~\ref{fig:first}(d)). We characterize the longitudinal dispersion of colloids as $\sigma^2_{||} = {\iint (x-\bar{x})^2 n ~dx dy}/{\iint n ~dx dy}$, where $\bar{x} = {\iint xn ~ dx dy}/{\iint n ~ dx dy}$ is the center of mass of the blob, which demonstrates an enhanced/suppressed dispersion in the attractive/repulsive cases, respectively (Fig.~\ref{fig:first}(e)). We non-dimensionalize these length scales using the lattice spacing $\lambda$. The transverse averaged light intensity further demonstrates the splitting in the attractive case and suppressed dispersion in the repulsive case (Fig.~\ref{fig:first}(f)). To probe the mechanism behind these observations we resort to numerical simulations.\\

\begin{figure}
\centering
    \includegraphics[width=\textwidth, keepaspectratio]{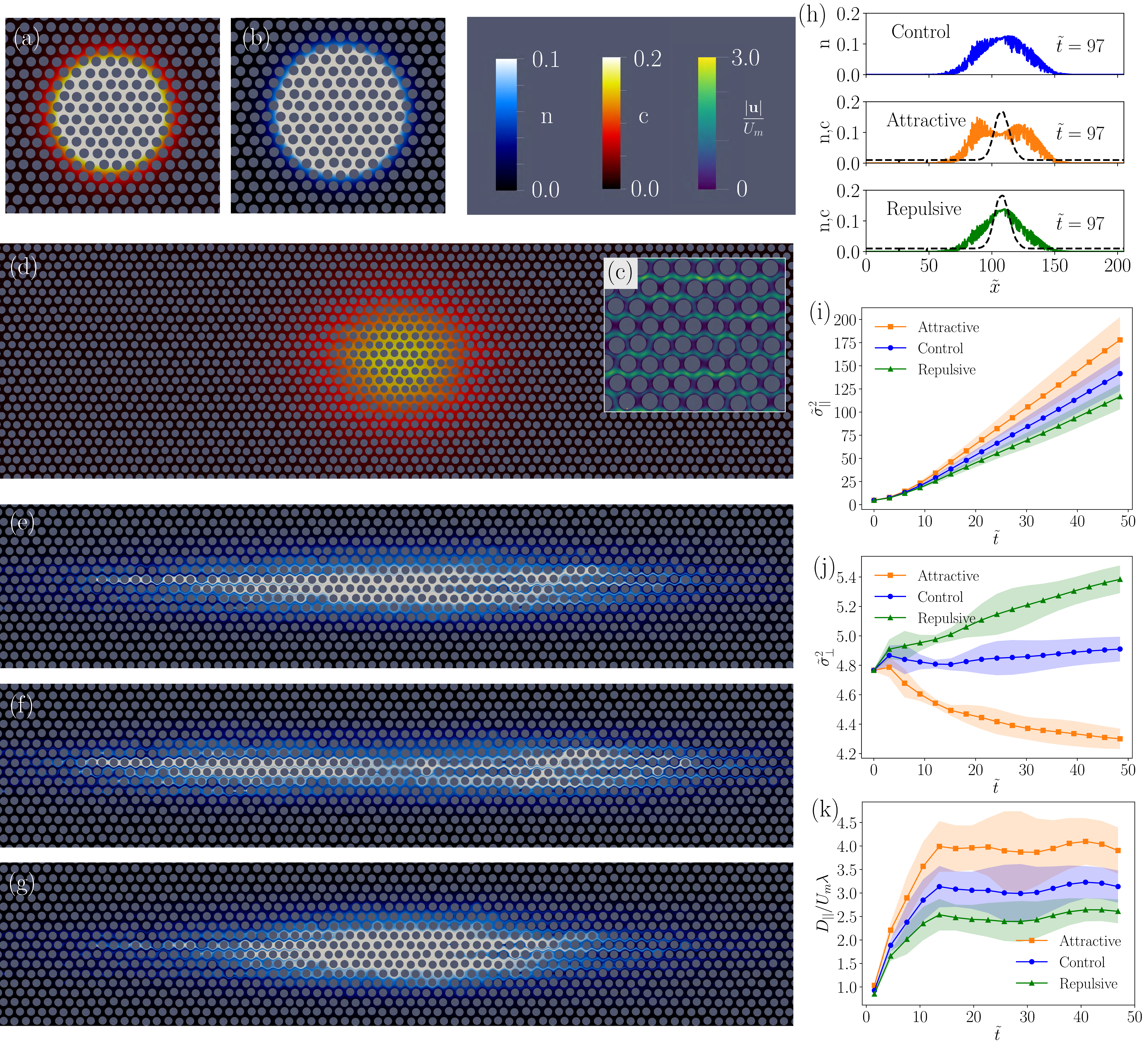}
    \captionsetup{width=\textwidth, justification=justified, singlelinecheck=false}  
\caption{ \label{fig:expt_vs_simu} 2D numerical simulations of the evolution of solute and colloid blobs. (a,b) A Gaussian blob of solute and colloid is introduced into the medium at $\tilde{t}=0$. Flow direction is from left to right with $Pe_c\approx 7200$ and $\chi=0.2$. Color scale is truncated for visual clarity. (c) The flow velocity distribution in the medium.  (d,e,f,g) Snapshots of the solute, and control, attractive, and repulsive cases of colloid fields at $\tilde{t}=tU_m/\lambda=97$, respectively. (h) The corresponding cross-sectionally averaged concentration profiles show the bimodal splitting in the attractive case and inhibited dispersion in the repulsive case. (i,j) The corresponding longitudinal and transverse dispersion of the blob. Error envelopes depict averaging over different realizations of disorder. (k) The longitudinal dispersion coefficient.} 
\end{figure}

\noindent \textbf{2D Numerical Simulations:} The fluid flow is governed by the Stokes equations:
\begin{align}
\label{eq: vel eqn}
    -\nabla p + \mu \nabla^2 \vec{u} & = 0 , \\ 
    \nabla \cdot \vec{u} & = 0,
\end{align}
where $\mathbf{u}\equiv (u,v)$, $p$ and $\mu$ are the fluid velocity, pressure and viscosity respectively. We introduce a joint 2D Gaussian blob of solute ($c$) and colloids ($n$) with characteristic length $l_b (> \lambda)$, i.e., $c(x,y,t=0)=0.99\exp({-(x^2+y^2)/2l_b^2})+0.01$ and $n(x,y,t=0)=\exp({-(x^2+y^2)/2l_b^2})$ respectively (Fig.~\ref{fig:expt_vs_simu}(a,b)). Both solute and colloids are then advected by the background flow $\mathbf{u}(x,y)$ (Fig.~\ref{fig:expt_vs_simu}(c)). The evolution of solute and colloid fields are governed by the advection-diffusion equations:
\begin{align}
\label{eq: c eqn}
    \partial_t c + \nabla \cdot (c\vec{u}) &= D_{\textrm{s}} \nabla^2 c, \\
\label{eq: n eqn}
    \partial_t n + \nabla \cdot \left[ n(\vec{u} + \vec{u}_{dp}) \right] &= D_c \nabla^2 n, 
\end{align}
where the diffusiophoretic velocity  $\mathbf{u}_{\textrm{dp}} = \Gamma_p \nabla \ln c$. To lower the computational costs, the Pe numbers in the simulations are a factor of 2.5 times smaller than those of experiments, i.e., $Pe_c \approx 7200$, $Pe_s \approx 12\,(\text{attractive}), 19\, (\text{repulsive})$. The 2D numerical simulations were performed using the finite volume method in OpenFOAM \citep{ault2018diffusiophoresis,alipour2026}. A triangular mesh of resolution $\sim2.5\mu m$ is used. Post boundaries and walls are treated with no slip boundary conditions. The diffusioosmotic mobility along PDMS surfaces is lower than the diffusiophoretic mobility of the colloidal particles. For glass surfaces, however, the two mobilities have been shown to be comparable in magnitude \cite{liu_diffusioosmotic_2025}.
Here, the effect of diffusioosmosis on the velocity profile is neglected in the simulations for simplicity and computational cost. The reasonable qualitative agreement between our 2D simulations and experiments (as will be shown) suggests that diffusioosmotic flows may influence the observations quantitatively, but not qualitatively.

While the solute blob has a Gaussian profile at long times ($t \approx 100 (\lambda / U_m) \approx 10 \lambda^2/D_s$), the colloid blob shows a longitudinal bimodal splitting in the attractive case (Fig.~\ref{fig:expt_vs_simu}(f)), and inhibited dispersion in the repulsive case (Fig.~\ref{fig:expt_vs_simu}(g)). The cross-sectionally averaged colloid concentration profiles of the simulations (Fig~\ref{fig:expt_vs_simu}(h)), and their longitudinal dispersion (Fig.~\ref{fig:expt_vs_simu}(i))  are also in agreement with those observed in the experiments. The transverse dispersion of the blob is slightly enhanced in the repulsive case and slightly suppressed in the attractive case (Fig.~\ref{fig:expt_vs_simu}(j)), but marginal overall in comparison to the longitudinal dispersion. The longitudinal dispersion coefficient $D_{||} = d\sigma^2_{x}/dt$ evolves at early times as the blob samples the velocity heterogeneities of the medium and solute gradients, before reaching a plateau at longer times (Fig.~\ref{fig:expt_vs_simu}(k)), showing that the attractive case disperses more than the control, while the repulsive case disperses less (Fig.~\ref{fig:expt_vs_simu}(k)). The increase of longitudinal dispersion in the attractive case is more pronounced in the simulations than in the experiments. We attribute this to: (i) the 2.5D nature of the experiments, i.e., the presence of top/bottom walls with no slip boundary conditions, and (ii) the asymmetry of the attractive and repulsive cases. In the repulsive case, colloids are pushed away from the low velocity zones. In these zones, phoresis dominates over advection, and therefore has a strong footprint. In contrast, in the attractive case, colloids are being pulled from the high velocity zones toward the low velocity zones. As the colloids are in the high velocity zones, their transverse motion due to diffusiophoresis will be much smaller than their longitudinal motion due to flow. Therefore, it will take a longer time until the difference between attractive and control case manifests itself.

\begin{figure}
\centering
    \includegraphics[width=\textwidth, keepaspectratio]{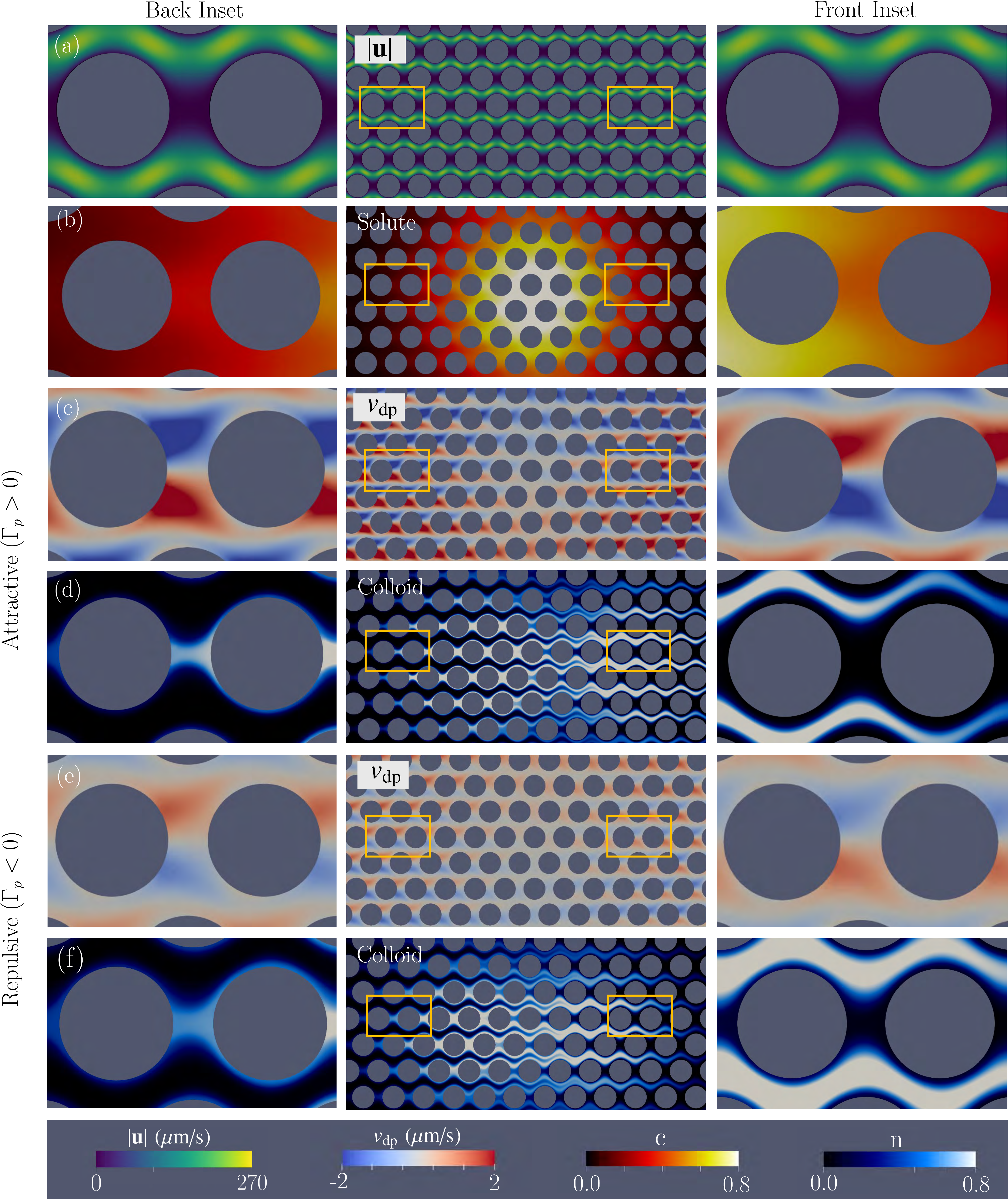}
    \captionsetup{width=\textwidth, justification=justified, singlelinecheck=false}  
\caption{\label{fig:pore_scale} Phoretic migration between slow and fast flow zones leads to macroscopic changes in the longitudinal dispersion of the blob. (a) The magnitude of the advective velocity $|\vec{u}|$ depicting fast zones in the channels and slow zones between channels. (b) As the blob is displaced, higher solute concentration pockets emerge in the low/high velocity zones in the back/front of the blob. (c) In the attractive case, transverse component of the phoretic velocity ($v_{\textrm{dp}}$) points toward the low/high velocity zones in the back/front of the blob, leading to the splitting of the blob in (d). (e) In the repulsive case, transverse component of the phoretic velocity points toward the high/low velocity zones in the back/front of the blob, leading to a less dispersed blob in the longitudinal direction (f). }
\end{figure}

To gain a better understanding of the mechanism behind our observations, we probe the early-time evolution of the blob. The origin of small scale transverse solute gradients is in the presence of shear, i.e., velocity gradient between slow and fast streamlines (Fig.~\ref{fig:pore_scale}(a)). In the attractive case, near the front of the solute blob concentration gradients point toward the high velocity pathways while pointing toward the low velocity zones near the back of the blob (Fig.~\ref{fig:pore_scale}(b)). These gradients lead to the focusing of the colloids in the high velocity zones near the front, and in the low velocity zones near the back of the blob, leading to enhanced dispersion and splitting of the colloidal blob (Fig.~\ref{fig:pore_scale}(c)). The phoretic migration is reversed in the repulsive case (Fig.~\ref{fig:pore_scale}(d,e)), leading to an enhanced transverse dispersion near the front, and removing the colloids from the low velocity zones near the back, and a longitudinally less dispersed colloidal blob. Effectively, in the attractive case, diffusiophoresis acts against diffusion, lowering the transverse dispersion, while in the repulsive case, diffusiophoresis works with diffusion, enhancing the transverse dispersion. Phoretic transport that occurs at the scale of the pores in a porous medium is similar to the channel-wide phoretic transport between fast and slow streamlines in a channel, as studied in the configuration of a 1D Gaussian pulse dispersing in Poiseuille flow by \cite{migacz2022diffusiophoresis}. We may relate the transverse solute gradients with the longitudinal gradient as: $dc/dy \sim Pe_s (dc/dx)$; therefore, as $Pe_s$ number decreases, the transverse solute gradients and the corresponding diffusiophoretic migration weaken. This explains the quantitative difference between experiments and the simulations at artificially small $Pe_s$ (Fig.~\ref{fig:first}(e) and Fig.~\ref{fig:expt_vs_simu}(i)). Further, since the solute blob remains Gaussian with its peak concentration decaying as $\sim 1/(D_{||,s}t/l_b^2)^{1/2}$, where $D_{||,s}$ is the longitudinal dispersion coefficient of the solute blob; therefore, the solute gradients decay slowly as $\sim t^{-1/2}$. Even though at long times $t\gg l_b^2/D_{||,s}$, we expect diffusiophoretic drift to become negligible, before this time, significant redistribution of the colloid has occurred already. Since  $D_c \ll D_s$, the particles do not diffuse back into their `control case configuration' even after solute gradients vanish, and therefore the disparity persists at long times.

\begin{figure}
\centering
    \includegraphics[width=\textwidth, keepaspectratio]{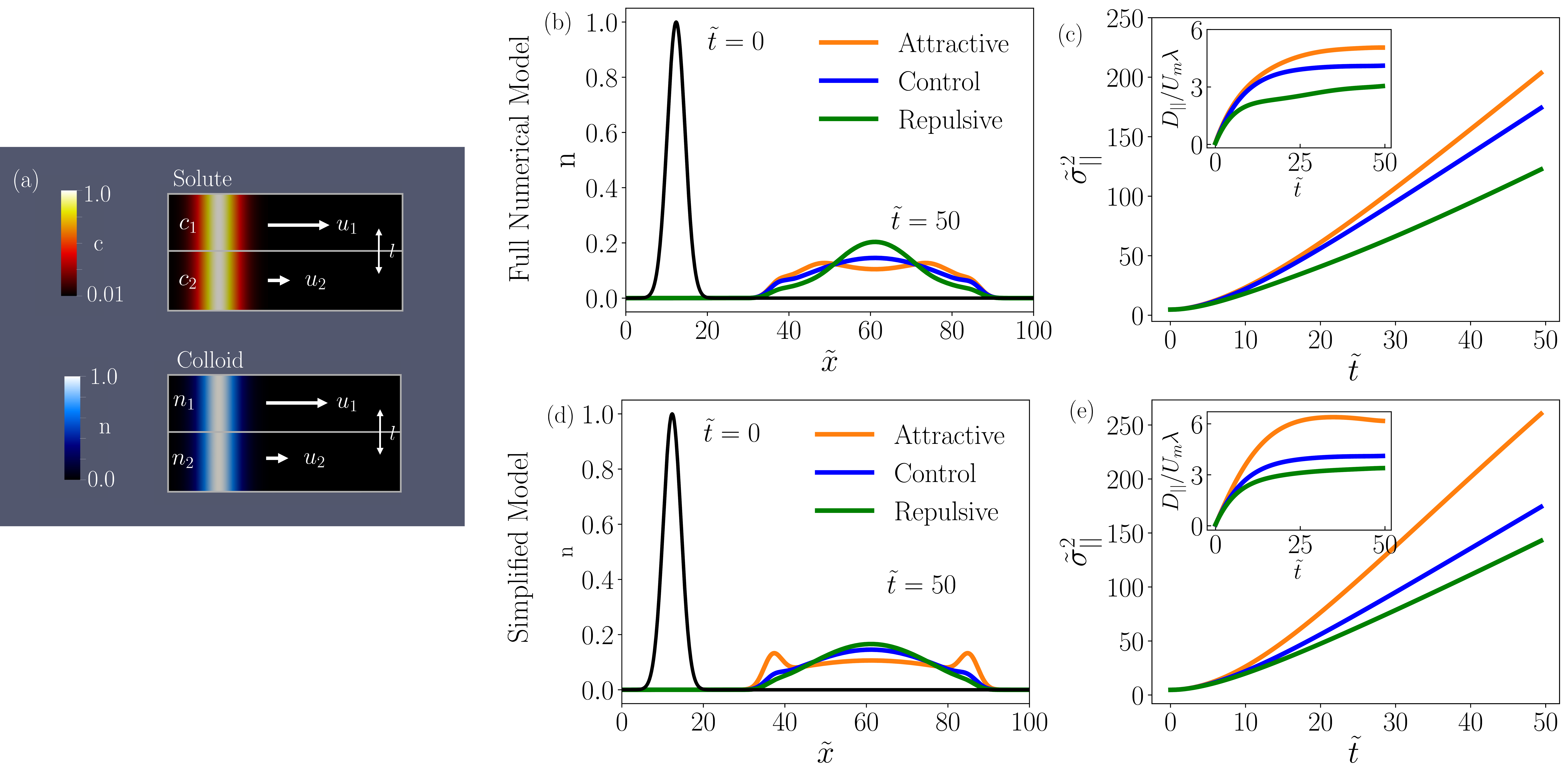}
    \captionsetup{width=\textwidth, justification=justified, singlelinecheck=false}  
\caption{\label{fig:2layer_model} (a) We construct a two-layer model with $u_1>u_2$ to probe the evolution of a 1D blob of solute and colloids. The two layers can communicate via diffusion and diffusiophoresis. (b) Full numerical model (Eqs.~\eqref{eq:solute} and \eqref{eq:colloid}) depicting the evolution of colloid density field $n=(n_1+n_2)/2$ for $Pe_c\sim7200$. The blob exhibits bimodal splitting in attractive case and inhibited dispersion in the repulsive case.  (c) The time evolution of second moment of the colloid blob ($\tilde{\sigma}^2_{||} = \sigma^2_{||}/l^2$) for all three cases and the corresponding dispersion coefficient (inset) show reasonable agreement with 2D numerical simulations and experiments. (d,e) Corresponding colloid density field, second moment, and dispersion coefficient (inset) obtained from the simplified model (Eqs.~\eqref{eq:solute} and \eqref{eq:simplified}). }
\end{figure}

\noindent \textbf{A minimal 2-layer model:} Using the insight provided by 2D simulations, we develop a 2-layer model with minimal ingredients to capture the main observations. While simplified models for dispersion in disordered media have been developed over the years \citep{bouchaud1990anomalous, guyon2012disorder}, here we use a two-layer model, similar in spirit to generalized Taylor dispersion \citep{van1990taylor}, to provide a mechanistic comparison with diffusiophoretic dispersion in Poiseuille flow. We consider two parallel layers with velocities $u_1$ and $u_2<u_1$ (Fig.~\ref{fig:2layer_model} (a)), and solute and colloid fields $c_i$, $n_i$, where $i=1,2$. The characteristic length scale between the two layers is $l$, and is chosen to be the pore width in the porous medium. We introduce a 1D Gaussian blob of solute and colloids, and observe how they evolve due to the background 1D flow, longitudinal and transverse diffusion and diffusiophoresis. In the frame moving with mean flow velocity $\bar{u}$, the governing equation for the solute field is:

\begin{align}
    \label{eq:solute}
    \partial_t \underbrace{\begin{bmatrix} c_1 \\ c_2 \end{bmatrix}}_{\vec{c}} + \underbrace{\begin{bmatrix} u_1 -\bar{u} & 0 \\ 0 & u_2 -\bar{u} \end{bmatrix}}_{\vec{U}} \partial_x \begin{bmatrix} c_1 \\ c_2 \end{bmatrix} 
    - D_{\textrm{s}} \partial_x^2 \begin{bmatrix} c_1 \\ c_2 \end{bmatrix} + & \underbrace{\frac{D_{\perp,s}}{l^2}\overbrace{\begin{bmatrix} 1 & -1 \\ -1 & 1 \end{bmatrix}}^\mathbb{A}}_{\vec{D}_{\perp,s}} \begin{bmatrix} c_1 \\ c_2 \end{bmatrix} = 0, \\
    \partial_t \underbrace{\begin{bmatrix} n_1 \\ n_2 \end{bmatrix}}_{\vec{n}} + \begin{bmatrix} u_1-\bar{u} & 0 \\ 0 & u_2-\bar{u} \end{bmatrix} \partial_x \begin{bmatrix} n_1 \\ n_2 \end{bmatrix} 
    - D_c \partial_x^2 \begin{bmatrix} n_1 \\ n_2 \end{bmatrix} + & \underbrace{\frac{D_{\perp,c}}{l^2}\begin{bmatrix} 1 & -1 \\ -1 & 1 \end{bmatrix}}_{\vec{D}_{\perp,c}} \begin{bmatrix} n_1 \\ n_2 \end{bmatrix} \notag  \\ 
    - \underbrace{\frac{v_{\textrm{dp}}}{l} \begin{bmatrix} H(-v_{\textrm{dp}}) & H(v_{\textrm{dp}}) \\ -H(-v_{\textrm{dp}}) & -H(v_{\textrm{dp}}) \end{bmatrix}}_{\vec{D}_{\perp,\textrm{dp}}} \begin{bmatrix} n_1 \\ n_2 \end{bmatrix} = 0, \label{eq:colloid}
\end{align}
where the tensors $\vec{D}_{\perp,s}$, $\vec{D}_{\perp,c}$, and $\vec{D}_{\perp,\textrm{dp}}$ represent the transverse dispersion of solute and colloids, and transverse phoretic exchange, respectively. $H(x)$ is the step function. Inter-layer transport by diffusiophoretic velocity is given by $v_{dp}\partial_yn_i \sim v_{dp}n_i/l$, where, $v_{dp} = \Gamma_p \frac{\partial \log{\text{(solute conc.)}}}{\partial y} \approx \Gamma_p\frac{(\log{c_1}-\log{c_2})}{l} = \frac{\Gamma_p}{l} \log{\frac{c_1}{c_2}}$. Here we choose the length scale while approximating the spatial derivative to be $l$. We assume $u_{\textrm{dp}} \ll u_1, u_2$, and is thus negligible. The initial conditions are: $c_{i}(x,0) = (c_0-c_\infty)e^{-x^2/(2l_b^2)} + c_\infty$, where $c_0=1, ~ c_\infty=0.01$, and $n_{i}(x,0) = n_0 e^{-x^2/(2l_b^2)}$, where $n_0=1$, and $l_b$ is the characteristic blob size.

While it is clear that the diffusiophoretic term $\vec{D}_{\perp,\textrm{dp}}$ in Eq.~(2.6) is not a diffusive term, we can gain insight by approximating it as one \citep{raynal2018advection}. Our 2D simulations showed that diffusiophoresis acts against the transverse dispersion of colloids in the attractive case, and acts together with it in the repulsive case. Thus, we can simplify $(\vec{D}_{\perp,c}+\vec{D}_{\perp,\textrm{dp}})\cdot\vec{n} \approx \vec{D}^{\textrm{eff}}_{\perp,\textrm{att}}\cdot\vec{n}=(D^{\textrm{eff}}_{\perp,\textrm{att}}/l^2)\mathbb{A}\cdot\vec{n}=(D_{\perp,c}/l^2+D_{\perp,att}(t)/l^2)\mathbb{A}\cdot\vec{n}$, with $D^{\textrm{eff}}_{\perp,\textrm{att}}<D_{\perp,\textrm{c}}$ for the attractive case, and $\approx \vec{D}^{\textrm{eff}}_{\perp,\textrm{rep}}\cdot\vec{n}=(D^{\textrm{eff}}_{\perp,\textrm{rep}}/l^2)\mathbb{A}\cdot\vec{n}=(D_{\perp,c}/l^2+D_{\perp,\textrm{rep}}(t)/l^2)\mathbb{A}\cdot\vec{n}$, with $D^{\textrm{eff}}_{\perp,\textrm{rep}}>D_{\perp,\textrm{c}}$ for the repulsive case. Here, we take $D_{\perp,\textrm{att/rep}}(t)=0.5\Gamma_p\max[\log{c_1(t)/c_2(t)}]$. Therefore, we expect the diffusiophoretic colloids to display the similar behavior as would a solute with a modified transverse dispersion at early times: $D^{\textrm{eff}}_{\perp,\textrm{att}}<D_{\perp,\textrm{c}}<D^{\textrm{eff}}_{\perp,\textrm{rep}}$. We note that as $t\rightarrow\infty$, $D^{\textrm{eff}}_{\perp,\textrm{att}}\rightarrow D_{\perp,c}$ and $D^{\textrm{eff}}_{\perp,\textrm{rep}}\rightarrow D_{\perp,c}$.

We proceed to analyze the simplified 2-layer transport equation with a transverse dispersion coefficient of $D_{\perp,j}$, dropping the superscript ``eff" for brevity: 
\begin{align}
\label{eq:simplified}
    \partial_t \vec{n} + \vec{U}\cdot\partial_x\vec{n} - D_c\partial_x^2\vec{n} + (D_{\perp,j}/l^2)\mathbb{A}\cdot\vec{n} = 0, 
\end{align}
where $j=\textrm{att}, \textrm{c}, \textrm{rep}$. Taking the Fourier transform of this equation, we then obtain the eigenmodes of the system, and analyze it in the diffusion- and advection-dominated regimes. On Fourier transformation we obtain, $ \partial_t \hat{\vec{n}} = \left( -i\vec{U} k -k^2 D_c \vec{I} - (D_{\perp,j}/l^2)\mathbb{A} \right) \cdot \hat{\vec{n}}$. Seeking solutions of the form $\hat{\vec{n}} = e^{\lambda_\pm t} \hat{\vec{n}}_{\lambda_{{\pm}}}$ yield the following two eigenvalues:
\begin{equation}
\label{eq: eigenvalues original}
    \lambda_{\pm} = -D_ck^2 - ik\bar{u} - D_{\perp,j}/l^2 \pm \beta(k),
\end{equation}
where $\beta(k) = \sqrt{(\bar{\bar{u}}^2 - \bar{u}^2)k^2 + D_{\perp,j}^2/l^4}$, and $\bar{u}$ and $\bar{\bar{u}}$ are the arithmetic and geometric mean (AM and GM) of $u_1$ and $u_2$ respectively. The AM-GM inequality implies that the coefficient of the $k^2$ term in $\beta(k)$ is necessarily negative, and $\beta(k)$ can become complex. The difference between the AM and GM of the velocities is indicative of shear in the 2-layer model, allowing us to define a ``shear parameter" $\gamma = \sqrt{\bar{u}^2 - \bar{\bar{u}}^2}$. We then have $\beta(k)=\sqrt{-\gamma^2k^2+D_{\perp,j}^2/l^4}$. In the limit when $\gamma k \ll D_{\perp,j}/l^2$, transverse dispersion dominates shear, leading to $\beta(k) \approx D_{\perp,j}/l^2 - \gamma^2l^2 k^2/2D_{\perp,j} $. The eigenvalues then are:
\begin{align}
\label{eq:lambda+ diffusion dominant}
    \lambda_+ & = -\left[ D_c + \gamma^2l^2/2D_{\perp,j} \right] k^2 - i\bar{u}k \\
\label{eq:lambda- diffusion dominant}
    \lambda_- & = -\left[  D_c - \gamma^2l^2/2D_{\perp,j} \right] k^2 - i\bar{u}k - 2D_{\perp,j}/l^2, 
\end{align}
where the first term corresponds to the dispersion, and the second term corresponds to advection. The constant term in Eq.~\eqref{eq:lambda- diffusion dominant} is a sink term. We find that both modes are advected with the same mean velocity $\bar{u}$. In the $\lambda_+$ mode, $D_{||}=D_c+\gamma^2l^2/2D_{\perp,j}$, i.e., an enhanced longitudinal dispersion. This mode persists as $t\rightarrow \infty$, while the $\lambda_-$ mode decays due to the sink term. In the opposite limit, where $\gamma k \gg D_{\perp,j}/l^2$, we have $\beta(k) \approx i (\gamma k - \frac{D_{\perp,j}^2}{2k\gamma l^4}) \approx i\gamma k$. The eigenvalues are:
\begin{align}
\label{eq:lambda+ shear}
    \lambda_{+} & = -D_ck^2 - i(\bar{u}-\gamma)k - D_{\perp,j}/l^2 \\ 
\label{eq:lambda- shear}
    \lambda_{-} & = -D_ck^2 - i(\bar{u}+\gamma)k - D_{\perp,j}/l^2,
\end{align}
 where the existence of two modes migrating at different velocities ($\bar{u}-\gamma$ and $\bar{u}+\gamma$) manifests as a ``splitting" of an initially unimodal distribution. Recalling that $D_{\perp,\textrm{att}}<D_{\perp,\textrm{c}}<D_{\perp,\textrm{rep}}$, we can gain some insight by assuming that the repulsive case corresponds to the diffusion-dominant regime $\gamma k \ll D_{\perp,\text{rep}}/l^2$, leading to $D_{||,\textrm{rep}}=D_c+\gamma^2l^2/2D_{\perp,\textrm{rep}} < D_c+\gamma^2l^2/2D_{\perp,c}$, i.e., suppressed longitudinal dispersion compared to the control case. In the opposite limit of shear-dominated transport, where $\gamma k \gg D_{\perp,\textrm{att}}/l^2$, we obtain the two modes traveling with different velocities, leading to the splitting, and enhanced dispersion.  
 
 \begin{figure}
\centering
    \includegraphics[width=\textwidth, keepaspectratio]{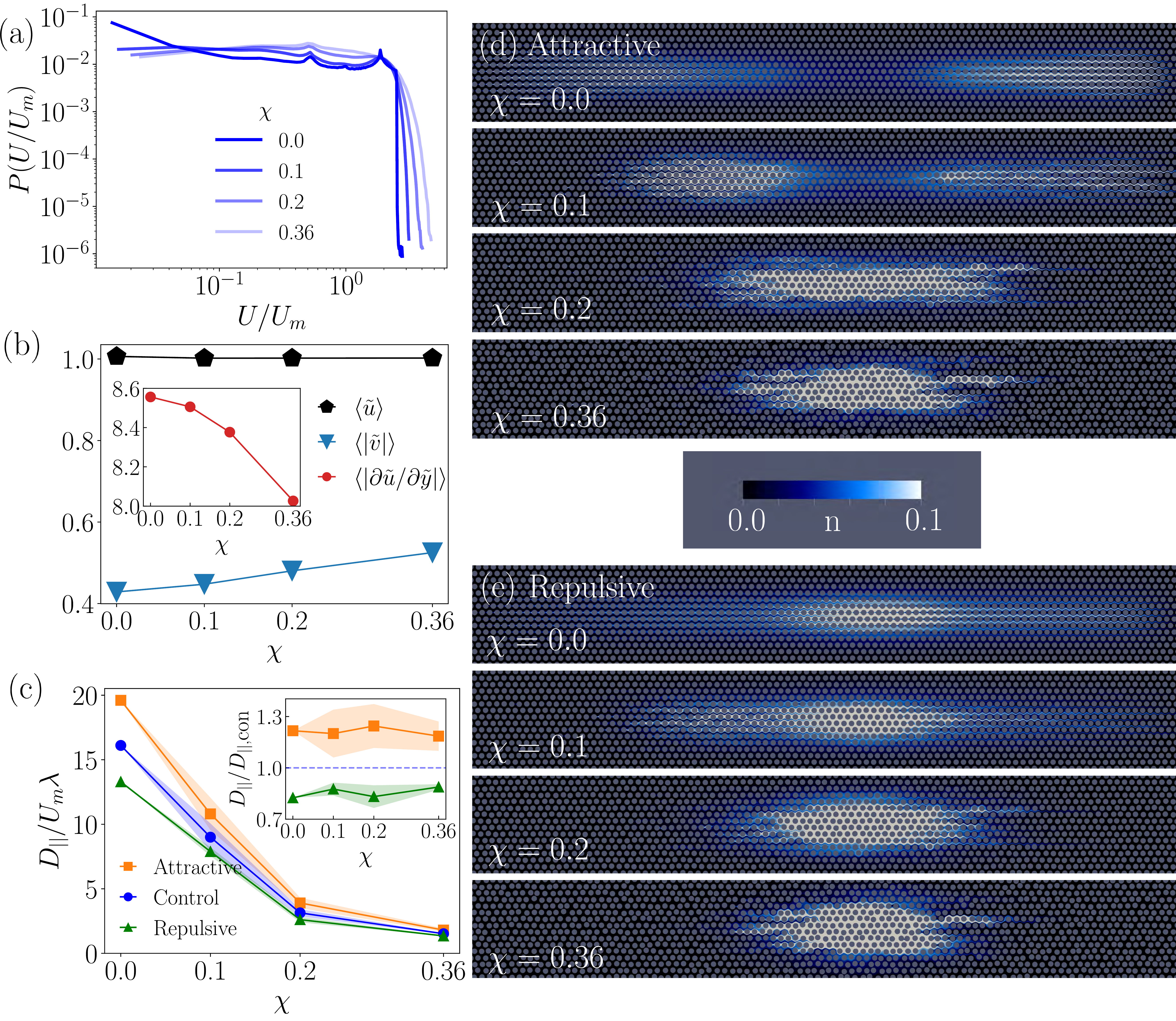}
    \captionsetup{width=\textwidth, justification=justified, singlelinecheck=false}  
\caption{ \label{fig:disorder_variation} The interplay of disorder and diffusiophoresis modulates the macroscopic dispersion of colloids. (a) Velocity distribution within the medium for different disorder strengths $\chi$. (b) The mean longitudinal velocity $\langle \tilde{u} \rangle $ is held constant throughout the simulations. The mean absolute transverse velocity $\langle |\tilde{v}| \rangle$ increases with disorder, while the mean absolute transverse shear $\langle |\partial \tilde{u} / \partial \tilde{y}| \rangle $ decreases (inset). (c) Dispersion coefficient versus disorder at time $\tilde{t}=48$ for $Pe_c \sim 7200$. Inset: Corresponding ratio of dispersion coefficient with respect to the control case. (d,e) Geometric disorder suppresses the influence of phoresis as the transverse velocity magnitude dominates phoretic exchange between fast and slow zones.}
\end{figure}

These results can also be interpreted in the broader context of chromatographic dispersion in porous media. In the simplified trapping model discussed by \citet{bouchaud1990anomalous}, a particle advected along a mobile backbone with velocity $V$ intermittently enters stagnant regions with probability $p$, acquiring a residence time $\tau$ over a characteristic longitudinal distance $\xi$. Using their notation, this gives an effective velocity $U^{-1}=V^{-1}+p\langle \tau\rangle/\xi$ and a longitudinal dispersion coefficient $D_{||}=pU^3(\langle \tau^2\rangle-p\langle \tau\rangle^2)/(2\xi)$. Equivalently, in the small-$p$ limit, this can be written as $D_{||}=\frac{1}{2}fU^2\langle \tau^2\rangle/\langle \tau\rangle$, where $f$ is the fraction of time spent in stagnant regions, showing that the dispersion scale is set by the ratio $\langle \tau^2\rangle/\langle \tau\rangle$. Thus, longitudinal dispersion is controlled by fluctuations in the time spent in different velocity states. Our two-layer model retains this same physical ingredient, but replaces mobile--immobile exchange by exchange between two mobile layers with different velocities and transverse exchange rate $\kappa=D_{\perp,j}/l^2$ (equivalently, an inter-layer residence time $\tau_\perp=\kappa^{-1}$). In the long-wavelength limit, the model gives $D_{||}=D_c+\gamma^2/(2\kappa)=D_c+\gamma^2\tau_\perp/2$, showing explicitly that reduced transverse exchange enhances longitudinal dispersion, whereas increased exchange homogenizes the blob and suppresses dispersion. In the present system, diffusiophoresis dynamically and sign-dependently modifies this exchange: attraction lowers the effective transverse exchange and, in the shear-dominated regime $\gamma k\gg \kappa$ identified above, can lead to bimodal longitudinal splitting, whereas repulsion increases the exchange and suppresses dispersion.

To verify these predictions, we numerically solve the coupled solute and colloid fields (Eqs.~\eqref{eq:solute}, \eqref{eq:colloid}), keeping the parameter values same as in the 2D simulations. The perpendicular dispersion coefficients are $D_{\perp,s}\sim D_s$ and $D_{\perp,c}\sim 10 D_c$ to reflect transverse dispersion due to geometric disorder. The layer thickness $l$ is chosen as the pore width for a hexagonal lattice given by $l=\lambda( 1-(2\sqrt{3}(1-\phi)/\pi)^{1/2} )$, which amounts to $l\approx\lambda/4$ for $\phi=0.5$. Velocities $u_1$ and $u_2$ are such that $u_1+u_2=2U_m$ and $u_1-u_2 \sim 0.5 \langle|\partial u/\partial y|\rangle l$. The mean shear rate is estimated as $\langle|\partial u/\partial y|\rangle \sim U_m/(l/2) \sim U_m/(\lambda/8)$.  The solution to the full numerical model (Eqs.~\eqref{eq:solute} and \eqref{eq:colloid}) is plotted as the average concentration of both layers $n \equiv (n_1+n_2)/2$ (Fig.~\ref{fig:2layer_model}(b)), along with that of the simplified model (Eqs.~\eqref{eq:solute} and \eqref{eq:simplified}) (Fig.~\ref{fig:2layer_model}(d)). Both the full numerical and simplified 2-layer models capture the longitudinal split of the colloid concentration in the attractive case, and inhibited dispersion with a unimodal distribution of the repulsive case (Fig.~\ref{fig:2layer_model}(a,e)). They also capture the observed qualitative trends for both longitudinal dispersion and dispersion coefficient in the presence of solute gradients (Fig.~\ref{fig:2layer_model}(c,e)), showing that the minimal necessary ingredients are shear and the transverse diffusiophoretic exchange of colloids. \\

\noindent \textbf{Impact of geometric disorder on colloid dispersion:} To alter the velocity distribution in the porous medium, we introduce disorder into the ordered hexagonal lattice and study four different configurations of increasing disorder via simulations: $\chi= 0.0 \, \, \text{(ordered)}, 0.1, 0.2, 0.36$ (Fig.~\ref{fig:disorder_variation}(a)), maintaining the mean flow velocity $\langle u \rangle$ constant (Fig.~\ref{fig:disorder_variation}(b)). As we increase the disorder, the mean transverse velocity magnitude, $\langle |v| \rangle$, increases (Fig.~\ref{fig:disorder_variation}(b)). The mean transverse shear, however, decreases with disorder (Fig.~\ref{fig:disorder_variation} (b) inset). These trends are indicative of transition from shear-dominated to diffusion-dominated transport, associated with enhanced transverse dispersion and suppressed longitudinal dispersion. In the attractive case, this transition implies the suppression of splitting, and in the repulsive case, a more compact blob (Fig.~\ref{fig:disorder_variation}(d,e)). While dispersion is suppressed with disorder (Fig.~5(c)), the relative impact of solute gradients on longitudinal dispersion remains constant (Fig.~\ref{fig:disorder_variation}(c) inset). Here, we have limited our study to weakly disordered media to avoid overlap of the posts. Many natural porous media, however, are strongly disordered, including regions of low permeability and dead-end zones, where diffusiophoretic transport is expected to dominate over advection \citep{park_microfluidic_2021,jotkar_impact_2024}. These stagnant zones facilitate colloid trapping while preferential flow pathways advect colloids; together, this allows the observation of bimodal dispersion of colloids to persist for strongly disordered media. Beyond dead-end pores, recent work by \citet{alipour2026} showed that diffusiophoretic migration of colloids within preferential flow pathways of porous media leads to significant changes of macroscopic dispersion of colloids. Therefore, we expect the influence of phoretic migration on dispersion of colloidal blobs to persist for highly disordered porous media.

\section{Summary and Conclusion}

Our work demonstrates that the interplay between geometric disorder and solute gradients modulates the macroscopic transport and dispersion of colloids in porous media. While anomalous, and even bimodal distributions have been observed in heterogeneous media, where multiple length scales are present \citep{koch1988anomalous,berkowitz_modeling_2006}, here we observed that even in a homogeneous porous medium, solute gradients can lead to bimodal/suppressed dispersion of colloids. We showed that the phoretic exchange between slow and fast flow zones leads to enhanced/reduced dispersion in the attractive/repulsive cases, contrary to the trends observed in phoretic transport of colloids in chaotic flows \citep{deseigne2014pinch,volk2014chaotic}. Incorporating these minimal ingredients in a 2-layer model, we could recover the trends observed in experiments and simulations. Together, our observations build on growing body of works on the importance of diffusiophoresis in transport of colloids in porous media \citep{park_microfluidic_2021,doan2021confinement,sambamoorthy_diffusiophoresis_2023,somasundar_diffusiophoretic_2023,jotkar_impact_2024,alipour2026}, highlighting the need for considering solutal effects at the pore-scale in macroscopic models of colloid transport.

\backsection[Funding.]{ We acknowledge the American Chemical Society Petroleum Research Fund (grant 65684-DNI9), Office of the Under Secretary of Defense for Research and Engineering (award FA9550-22-1-0320), and National Science Foundation (CAREER award 2443484) for the partial support of this work.}

\backsection[Declaration of interests]{The authors report no conflict of interest.}

%

\bibliographystyle{jfm}
\bibliography{jfm}


\end{document}